\documentclass[12pt,letterpaper,oneside]{article}
\usepackage[utf8]{inputenc}
\usepackage[english]{babel}
\usepackage{amsmath}
\usepackage{amsfonts}
\usepackage{amssymb}
\usepackage{graphicx}
\usepackage{aas_macros}
\usepackage{xcolor}
\usepackage[top=1in, left=0.9in, right=1.25in, bottom=1in]{geometry}
\usepackage{setspace}
\usepackage[outercaption]{sidecap}
\usepackage[backend=biber,style=authoryear,uniquename=false,maxcitenames=2,giveninits=false,natbib=true]{biblatex} 
\usepackage[autostyle=true]{csquotes} 
\DeclareLanguageMapping{spanish}{spanish-apa}
\defbibenvironment{bibliography}
  {\enumerate
     {}
     {\setlength{\leftmargin}{\bibhang}%
      \setlength{\itemindent}{-\leftmargin}%
      \setlength{\itemsep}{\bibitemsep}%
      \setlength{\parsep}{\bibparsep}}}
  {\endenumerate}
  {\item}
\DefineBibliographyStrings{english}{andothers = {et\addabbrvspace al\adddot}, and={\&}}

\begin{document}
\begin{center}
Proceedings IAU Symposium No. 376. At the cross-roads of astrophysics and cosmology: Period-luminosity relations in the 2020s.

\vspace{1.5cm}

{\LARGE Globular cluster metallicities and distances from}\\
\vspace{1.0cm}

{\LARGE disentangling their RR Lyrae light curves}
\vspace{1.5cm}

{\LARGE A. Arellano Ferro}

\vspace{0.9cm}
{\Large Instituto de Astronom\'ia.}\\
\vspace{0.5cm}
{\Large Universidad Nacional Autónoma de M\'exico}

\end{center}
\newpage
\spacing{0.935}

\begin{abstract}
We present mean horizontal branch absolute magnitudes and iron
abundances for a sample of 39 globular clusters. These quantities were
calculated in an unprecedented homogeneous fashion based on Fourier
decomposition of ligt curves of RR Lyrae cluster members. Zero points
for the luminosity calibrations are discussed. Our photometrically
derived metallicities and distances compare very well with
spectroscopic determinations of [Fe/H] and accurate distances obtained
using {\sl Gaia} and {\sl Hubble Space Telescope} data. The need to
distinguish between the results for RRab and RRc stars for a correct
evaluation of the $M_V$--[Fe/H] relation is discussed. For RRab stars,
the relation is non-linear, and the horizontal branch structure plays
a significant role. For RRc stars, the relation remains linear and
tight, and the slope is very shallow. Hence, the RRc stars seem better
indicators of the parental cluster distances. Systematic time-series
CCD imaging performed over the last 20 years enabled to discover and
classify 330 variables in our sample of globular clusters.

\end{abstract}



\section{Introduction}

The Fourier light curve decomposition technique is based on the
representation of the observed light curve by a series of harmonics of
adequate amplitudes and displacements, $A_k$ and $\phi_k$, in an
equation of the form,
\begin{equation}
\label{eq.Foufit}
m(t) = A_0 + \sum_{k=1}^{N}{A_k \cos\left({2\pi k\over P}~(t-E) + \phi_k\right) },
\end{equation}
where $m(t)$ is the magnitude at time $t$, $P$ is the pulsation
period, and $E$ is the epoch, typically selected as the time of
maximum brightness. A linear minimization routine is used to derive
the best-fitting values for the amplitudes $A_k$ and phases $\phi_k$
of the sinusoidal components. The Fourier parameters, defined as
$\phi_{ij} = j\phi_{i} - i\phi_{j}$ and $R_{ij} = A_{i}/A_{j}$, are
computed from the amplitudes and phases of the harmonics in
eq.~\ref{eq.Foufit}. In principle, a proper combination of some of
these Fourier parameters, to be determined on theoretical and/or
empirical grounds, can yield appropriate physical stellar
parameters. The demonstration of this fact and its practical
application has taken a few decades, and it traces its origins to the
pioneering work of \citet{vanAlbada1971}, \citet{Simon1988a} and \citet{Simon1988b}.
Simon employed hydrodynamic pulsation models to construct light curves
and experimented with Fourier decomposition in search of useful
correlations, eventually expressing that ``Finally, we argue that if a
reliable hydrodynamic pulsation code were available, the Fourier
technique would be capable of determining masses, luminosities and
temperatures of individual stars, in many cases using observations
already in hand.'' And, in fact, hydrodynamical models were developed
in another early work by \citet{Simon1993}, where the significance of
the $\phi_{31}$ parameter in the estimation of the stellar mass and
luminosity of RRc stars was demonstrated.

Towards the end of the 20th century, a program aimed at calculating
accurate empirical relations for the physical parameters of RR Lyrae
stars was developed at the Konkoly Observatory. Soon, robust and
simple empirical calibrations emerged for the determination of [Fe/H]
\citep{Jurcsik1996} and the absolute magnitude \citep{Kovacs1998} in terms
of the pulsation period, yielding the Fourier $A_1$ and $\phi_{31}$
parameters for a large sample of RRab stars in the globular clusters
of the Sculptor dwarf galaxy. The then-recent set of iron abundances
\citep{Suntzeff1994,Layden94} and photometric data
\citep{KovZso95,Lub77} were employed.

Soon after, it was argued that some hydrodynamical pulsation models
did not follow the empirical light curves \citep{Kovacs1998,KoKan1998},
and then empirical calibrations were preferred \citep{Kovacs2001}.

A thorough [Fe/H] calibration for RRc stars \citep{Morgan2007} employed
a sample of 106 calibrator stars in globular clusters and rendered
useful formulae for the iron abundance estimation on the Zinn--West
metallicity scale \citep{Zinn1984}. More recently, a further effort
to improve the accuracy of the calibrations, using high precision {\sl
  Kepler} photometry and detailed spectroscopic iron abundances for
RRab and RRc stars, was performed \citep{Nemec2013}. Twenty-six
calibrators were included in the RRab calibration (nine with Blazhko
modulations) and 101 calibrators for the RRc calibration.

For nearly 20 years our research group has been interested in
performing extensive time-series CCD photometry in the \emph{VI} bands
for a sample of Galactic globular clusters. We aimed at estimating
cluster mean metallicities and distances from their member RR Lyrae
stars via the Fourier decomposition technique, and from other
variables in a cluster when available, e.g. SX Phe via their
period--luminosity (P--L) relation, or red giant semi-periodic
variables near the tip of the red giant branch (RGB). Meanwhile, other
aspects of the clusters under study have been addressed, such as their
reddening, stellar membership and a census of their variable star
populations; their periods and classifications have also been
updated. As a natural by-product we discovered and classified about
330 new variables.

Numerous RR Lyrae light curve decompositions and their implied
physical parameters can be found in the literature, but often
different calibrations and zero points to estimate [Fe/H] and $M_V$
are used. For the determination of these quantities, in earlier
investigations, our group also used some different equations, since we
were also learning what were probably the best and most reliable
calibrations. Hence, we decided to reconsider all our data and
determine Fourier fits to pursue homogeneous calculations with the
selected calibrations, for our entire sample of globular clusters
\citep{Arellano2022}. The specific calibrations employed will be presented in
Section \ref{calibrations}.

The present paper is organized as follows. In Section
\ref{FeMvresults}, we summarize the resulting values of reddening,
[Fe/H] and $M_V$ (and, hence, distance) for our sample of globular
clusters and compare them with spectroscopically determined iron
abundance values, while comparing the distances with well established
mean distances based on {\sl Gaia} and {\sl Hubble Space Telescope}
({\sl HST}) data. In Section \ref {MvFeRel} we will revisit the
$M_V$--[Fe/H] correlation in light of the homogeneous set of Fourier
[Fe/H] and $M_V$ values. We highlight the need to split the discussion
of results obtained from RRab stars and those from RRc stars, as they
are of an obviously different nature. In Section \ref{HBparam} we
introduce the role of the HB structure on the $M_V$--[Fe/H] relation
and discuss its results. Section \ref{VarsGC} recounts briefly the
census of new variables discovered in this project. Finally, in
Section \ref{Summary} we summarize our results and conclusions.
  
\section{CCD Photometry based on the DIA approach}
\label{DIA}

All our CCD photometry obtained over the years has been performed
based on Difference Image Analysis (DIA) through the pipeline DanDIA
\citep{Bramich2008}; \citep{Bramich2013}; \citep{Bramich2015}. A detailed description of the
method and its caveats can be found in \citet{Bramich2011}.  A good
overview of all our procedures applied to the analysis of the
individual clusters is found in \citet{Deras2019} and \citet{Yepez20}.

\section{The [Fe/H] and $M_V$ calibrations employed}
\label{calibrations}

\subsection{The [Fe/H] calibrations}

For the calculation of [Fe/H] we adopted the following calibrations:
\begin{eqnarray}
\label{eq:RR0_Fe}
{\rm [Fe/H]}_{\rm J}= - 5.038 - 5.394 P + 1.345 \phi^{(s)}_{31},
\end{eqnarray}
and

\begin{eqnarray}
\label{eq:RR1_Fe}
{\rm [Fe/H]}_{\rm ZW} = 52.466 P^2 - 30.075 P + 0.131 \phi^{(c)2}_{31} - 0.982 \phi^{(c)}_{31} - 4.198 \phi^{(c)}_{31} P + 2.424,
\end{eqnarray}
from \citet{Jurcsik1996} and \citet{Morgan2007} for RRab and RRc stars,
respectively. The phases $\phi^{(c)}$ and $\phi^{(s)}$ are calculated
either based on a cosine or a sine series, respectively, and they are
correlated via $\phi^{(s)} = \phi^{(c)} - \pi$. The iron abundance on
the \citet{Jurcsik1996} scale, [Fe/H]$_{\rm J}$, can be converted to the
\citet{Zinn1984} scale via [Fe/H]$_{\rm J}$ = 1.431 [Fe/H]$_{\rm ZW}$
+ 0.88 \citep{Jurcsik1995}. All [Fe/H]$_{\rm ZW}$ values can be
converted to the UVES spectroscopic scale of \citet{Carretta2009} via
[Fe/H]$_{\rm UV} = -0.413 + 0.130$~[Fe/H]$_{\rm ZW} -
0.356$~[Fe/H]$_{\rm ZW}^2$.

We have also considered the more recent non-linear calibrations
calculated by \citet{Nemec2013} for both RRab and RRc stars. These
calibrations made use of high-precision {\sl Kepler} photometry and
spectroscopic iron abundances for 26 RRab and 101 RRc
calibrators. Their resulting equations are, respectively,

\begin{eqnarray}
{\rm[Fe/H]_{N}}=-8.65-40.12P+5.96\phi_{31}^{(s)}(K) +6.27\phi_{31}^{(s)}(K)P-0.72\phi_{31}^{(s)}(K){^2},
\label{NemAB}
\end{eqnarray}
where $\phi_{31}^{(s)}(K) = \phi_{31}^{(s)}+0.151$ is given on the
{\sl Kepler} scale \citep{Nemec2013}, and

\begin{eqnarray}
{\rm [Fe/H]_{N}}=1.70-15.67P+0.20\phi_{31}^{(c)}-2.41\phi_{31}^{(c)}P +18.0P^2  +0.17\phi_{31}^{(c)^2}.
\label{NemC}
\end{eqnarray}

The [Fe/H]$_{\rm N}$ values resulting from these calibrations,
calibrated with respect to spectroscopic iron abundances, should
produce iron abundances on the spectroscopic UVES scale, [Fe/H]$_{\rm
  UV}$.

\subsection{The $M_V$ calibrations}

For the calculation of $M_V$ we adopted the calibrations:

\begin{equation}
\label{eq.RR0_Mv}
M_V= -1.876~\log P -1.158 A_1 + 0.821 A_3 + 0.41,
\end{equation}

\begin{equation}
\label{eq.RR1_Mv}
M_V= -0.961 P - 0.044 \phi^{(s)}_{21} -4.447 A_4 + 1.061,
\end{equation}

\noindent
from \citet{Kovacs2001} and \citet{Kovacs1998} for the RRab and RRc stars,
respectively. We call attention to the zero points in the above two
equations, which do not correspond to those in the original
publications. The zero points of eqs. \ref {eq.RR0_Mv} and \ref
{eq.RR1_Mv} have been calculated to scale the luminosities of RRab and
RRc stars to the distance modulus of 18.5 mag for the Large Magellanic
Cloud (LMC) \citep{Freedman2001}; \citep{Pietrzynsky13} and \citep{deGrijs2014}, assuming a mean
magnitude for the RR Lyrae stars in the LMC of $V_0 = 19.064 \pm
0.064$ mag \citep{Clementini2003}, and a mean absolute magnitude for the
star RR Lyrae of $M_V = 0.61 \pm 0.10$ mag \citep{Benedict2002}. Other
discussions as to these zero points can be found in \cite{Kinman2002}
and \citet{Cacciari2005}. We refer the reader to section 4.2 of
\citet{Arellano2010} for a detailed discussion of the above zero point values
that we finally adopted.

\section{The Fourier {\rm [Fe/H]} and distance values for globular clusters}
\label{FeMvresults}

\begin{table*}
\tiny
\begin{center}
\caption{Mean values of [F\MakeLowercase{e}/H], on three different
  scales, and $M_V$ from a homogeneous Fourier decomposition of the
  light curves of cluster member RR Lyrae stars.$^{1,2}$}
\label{MV_FEH:tab}


\begin{tabular}{lc|ccccc|ccccc|cc}

\hline
GC &Oo& [Fe/H]$_{\rm ZW}$ & [Fe/H]$_{\rm UV}$& [Fe/H]$_{\rm N}$&$M_V$& N& [Fe/H]$_{\rm ZW}$ &[Fe/H]$_{\rm UV}$ &[Fe/H]$_{\rm N}$ & $M_V$ & N & $E(B-V)$&$HBt$\\
 & & dex & dex & dex &mag&  & dex &dex &dex & mag&   & mag& \\
\hline
NGC (M) & &&& RRab  & &&&& RRc& & &  &\\
\hline
1261     &I &--1.48$\pm$0.05&--1.38&--1.27&0.59$\pm$0.04 &6&--1.51$\pm$0.13 &--1.38 &--1.41&0.55$\pm$0.02  &4&0.01& --0.71\\

1851      &I &--1.44$\pm$0.10&--1.33&--1.18&0.54$\pm$0.03 &10&--1.40$\pm$0.13&--1.28 &--1.28 & 0.59$\pm$0.02& 5&0.02&--0.36\\

3201      &I &--1.49$\pm$0.10&--1.39&--1.29&0.60$\pm$0.04 & 19&--1.47$\pm$0.08&--1.37 &--1.36 & 0.58$\pm$0.01 &2&diff.&+0.08\\

4147      &I &--&--&--&-- & -- & --1.72$\pm$0.26&--1.68&--1.66 &0.57$\pm$0.05&6 & 0.01&+0.55\\

5272 (M3)  &I &--1.56$\pm$0.16&--1.46&--1.46&0.59$\pm$0.05 &59&--1.65$\pm$0.14&--1.57&--1.56 &0.56$\pm$0.06&23 &0.01&+0.08\\

5904 (M5)  &I &--1.44$\pm$0.09&--1.33&--1.19&0.57$\pm$0.08  &35&--1.49$\pm$0.11&--1.39&--1.38 &0.58$\pm$0.03&22 &0.03&+0.31\\

6171 (M107)&I &--1.33$\pm$0.12&--1.22&--0.98&0.62$\pm$0.04  & 6 &--1.02$\pm$0.18&--0.90&--0.88 &0.59$\pm$0.03&4 &0.33&--0.73\\

6229      &I &--1.42$\pm$0.07&--1.32&--1.13&0.61$\pm$0.06  & 12&--1.45$\pm$0.19&--1.32&--1.58&0.53$\pm$0.10&8&0.01&+0.24\\

6266$^7$ (M62) &I  &--1.31$\pm$0.11&--1.19& --  & 0.63$\pm$0.03 &40 &--1.23$\pm$0.09&--1.11& --  &0.51$\pm$0.03&21&diff&+0.55 \\

6362     &I &--1.25$\pm$0.06&--1.13&--0.83&0.62$\pm$0.01&2&--1.21$\pm$0.15&--1.09&--1.10 &0.59$\pm$0.05&6&0.06&--0.58 \\

6366     &I &--0.84&--0.77&--0.31&0.71  & 1 &-- &-- &--&-- & &0.80& --0.97\\

6401      &I &--1.36$\pm$0.09&--1.24&--1.04&0.60$\pm$0.07  &19 &--1.27$\pm$0.23&--1.09 &--1.16 &0.58$\pm$0.03&9&diff&+0.13\\

6712   &I &--1.25$\pm$0.06 &--1.13&--0.82&0.55$\pm$0.03&6 &--1.10$\pm$0.04&--0.95& --0.96&0.57$\pm$0.18 &3&0.35&--0.62\\

6934   &I &--1.56$\pm$0.14 &--1.48&--1.49&0.58$\pm$0.05&15 &--1.53$\pm$0.12&--1.41& --1.50&0.59$\pm$0.03 &5&0.10&+0.25\\

6981 (M72) &I &--1.48$\pm$0.11&--1.37&--1.28&0.63$\pm$0.02  &12 &--1.66$\pm$0.08&--1.60&--1.55 &0.57$\pm$0.04& 4&0.06&+0.14\\

7006  &I &--1.51$\pm$0.13&--1.40&--1.36&0.61$\pm$0.03  &31 &--1.53&--1.44&--1.43&0.55& 1&0.08&--0.28\\

Pal 2 & I &--1.39$\pm$0.55&--1.09&--1.20&0.52$\pm$0.08 &11&--&--& --&-- & &0.93 &-- \\

Pal 13  & I&--1.64$\pm$0.15 &--1.56 &--1.67 & 0.65$\pm$0.05 & 4 &--&--& --&-- & -- &0.10 &--0.30\\

\hline

288       & II&--1.64&--1.58&--1.42&  0.38& 1&--1.59 &--1.52 &--1.54 &0.58& 1 &0.03&+0.98\\

1904 (M79) & II&--1.63$\pm$0.14&--1.55&--1.47&  0.41$\pm$0.05& 5 &--1.71 &--1.66 &--1.69&0.58 & 1 &diff&+0.74\\

4590 (M68) & II&--2.07$\pm$0.09$^4$&--2.21&--2.01&  0.49$\pm$0.07& 5 &--2.09$\pm$0.03&--2.24&--2.23 &0.53$\pm$0.01&15 &0.05&+0.17 \\

5024 (M53) & II&--1.94$\pm$0.06$^4$&--2.00&--1.68&  0.45$\pm$0.05 &18&--1.84$\pm$0.13&--1.85&--1.85 &0.52$\pm$0.06& 3 &0.02&+0.81\\

5053      & II&--2.05$\pm$0.14$^4$&--2.18&--2.07&  0.46$\pm$0.08 & 3&--2.00$\pm$0.18&--2.05&--2.06 &0.55$\pm$0.05& 4&0.18&+0.50\\

5286$^7$ &II &--1.68$\pm$0.15&--1.64&--&0.52$\pm$0.04 &59&--1.71$\pm$0.23&--1.68&--&0.57$\pm$0.04&23 &0.24&+0.80\\

5466      & II&--2.04$\pm$0.14$^4$&--2.16&--2.01& 0.44$\pm$0.09 & 8 &--1.90$\pm$0.21 &--1.89&--1.96&0.53$\pm$0.06 & 5 &0.00&+0.58 \\

6139      &II& --1.63$\pm$0.02    &--1.57&--1.33& 0.49$\pm$0.06 & 2 &--1.62$\pm$0.39 &--1.55&--1.47&0.56$\pm$0.10 & 2 & diff& -- \\

6205 (M13)  & II&--1.60 &--1.54 &--1.00 & 0.38  & 1 &--1.70$\pm$0.20  &--1.63&--1.71 &0.59$\pm$0.05 & 3 &0.02&+0.97\\

6254 (M10)  & II?&-- &-- &--& --  & -- &--1.59&--1.52&--1.52 &0.52 & 1 &0.25&+1.00\\

6333 (M9)  & II&--1.91$\pm$0.13$^4$&--1.96&--1.72&  0.47$\pm$0.04& 7 &--1.71$\pm$0.23 &--1.66&--1.66 &0.55$\pm$0.04& 6 & diff&+0.87\\

6341 (M92) & II&--2.12$\pm$0.18$^4$&--2.16$^5$&--2.26&   0.45$\pm$0.03& 9&--2.01$\pm$0.11&--2.11&--2.17 &0.53$\pm$0.06& 3 &0.02&+0.91\\

6809$^7$  &II &--1.61$\pm$0.20&--1.55&--&0.53$\pm$0.09 &59&--&--&--&--&--&0.08&+0.87\\

7078 (M15) & II&--2.22$\pm$0.19$^4$&--2.46&--2.65&  0.51$\pm$0.04 &9&--2.10$\pm$0.07&--2.24&--2.27 &0.52$\pm$0.03 & 8 &0.08&+0.67 \\

7089 (M2)  & II&--1.60$\pm$0.18&--1.51&--1.25 & 0.53$\pm$0.13& 10 &--1.76$\pm$0.16&--1.73&--1.76 &0.51$\pm$0.08& 2&0.06&+0.38$^5$\\

7099 (M30) & II&--2.07$\pm$0.05$^4$&--2.21&--1.88&  0.40$\pm$0.04& 3&--2.03  &--2.14&--2.07 &0.54& 1 &0.03&+0.89\\

7492$^6$  & II&--1.89$^{4,5}$&--1.93&--0.83& 0.37 & 1&--&--&-- &-- &--&0.00&+0.76 \\

\hline

6402 (M14) &Int&--1.44$\pm$0.17 &--1.32&--1.17&0.53$\pm$0.07 &24&--1.23$\pm$0.21&--1.12&--1.12&0.58$\pm$0.05&36&0.57&+0.65 \\

6779 (M56) &Int&--1.97$^4$&--2.05&--1.74&0.53 &1 &--1.96&--2.03&--2.05&0.51&1&0.26&+0.98\\
\hline
6388  &III&--1.35$\pm$0.05&--1.23&--1.00&0.53$\pm$0.04&2&--0.67$\pm$0.24&--0.64&--0.56 &0.61$\pm$0.07&6&0.40&--1.00 \\
6441  &III&--1.35$\pm$0.17&--1.23&--0.80&0.43$\pm$0.08&7&--1.02$\pm$0.34&--0.82&--1.00 &0.55$\pm$0.08&8&0.51&--0.73\\

\hline

\end{tabular}
\small
\center{\quad Notes: $^1$ Original light curves and Fourier parameters
  can be found in the papers long list in table 1 of \citet{Arellano2022}. {\bf
    Unfortunately the table 1 as published in \citet{Arellano2022} was an older
    version and exhibits some small differences compared with the
    present table, which in fact is used for the production of all
    figures and discussion in the present paper. The present table
    supersedes that in \citet{Arellano2022} } $^2$ Quoted uncertainties are
  1$\sigma$ errors calculated from the scatter in the data for each
  cluster. The number of stars considered in the calculations is given
  by $N$. $^3$ The only RRL V1 is probably not a cluster member. $^4$
  This value has --0.21 dex added; see \citet{Arellano2022} for a
  discussion. $^5$ Our calculation. $^6$ Based on a single light curve
  that was not fully covered. $^7$ Metallicity and $M_V$ taken from
  the compilation of \citet{Contreras2010}.\\ }
\end{center}
\end{table*}

\begin{figure}
\begin{center}
\includegraphics[scale=.7]{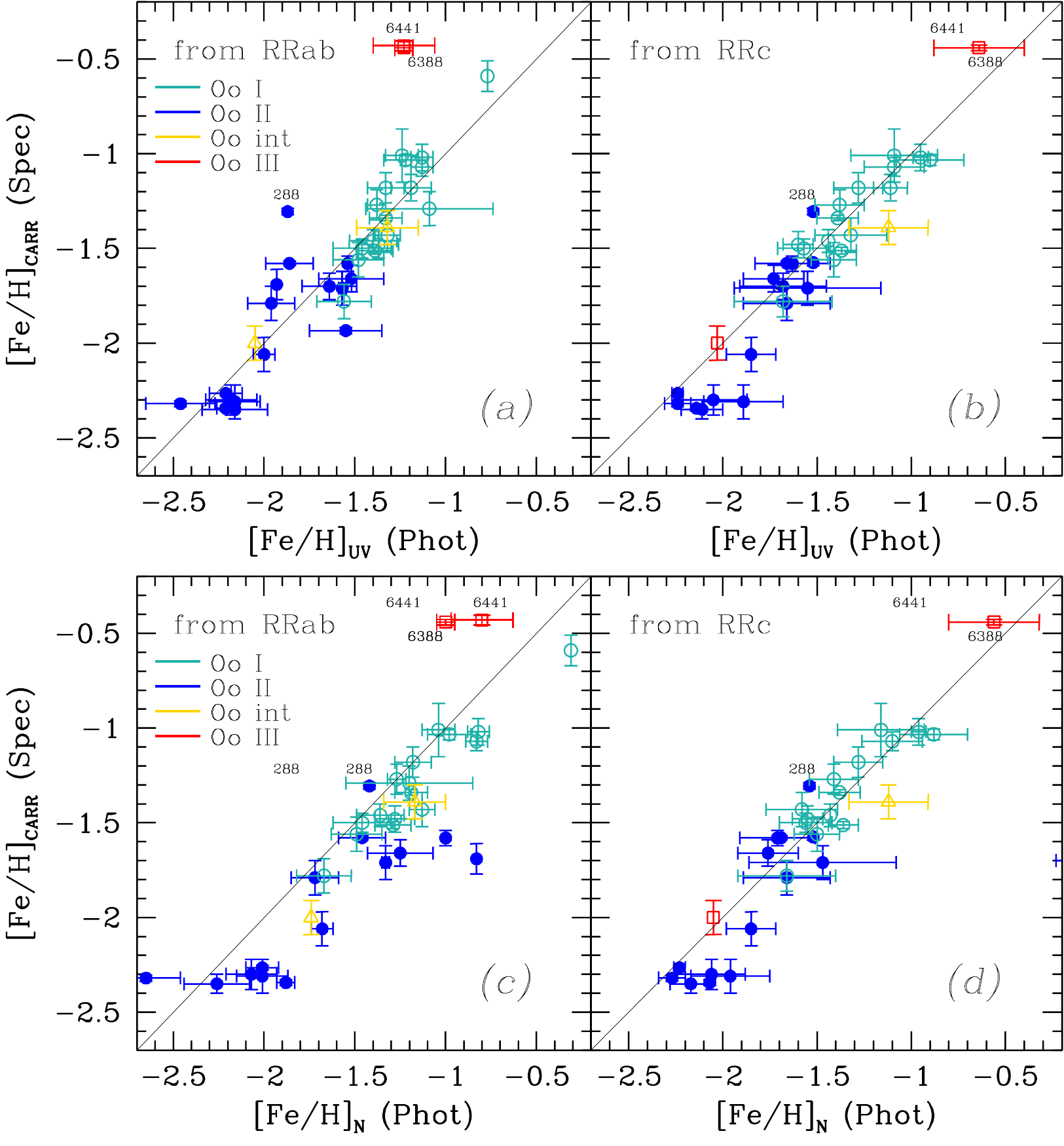}
\caption{Comparison of the photometric, Fourier decomposition-based
  [Fe/H]$_{\rm UV}$ and [Fe/H]$_{\rm N}$ (Table \ref{MV_FEH:tab}),
  with equivalent values from high-resolution spectroscopy,
  [Fe/H]$_{\rm Carr}$ \citep{Carretta2009}. (Updated from figure 1 in
  Arellano Ferro 2022, Rev. Mex. A\&A, 58, 257). The color figure can
  be viewed online.}
\label{PhoVSspec}
\end{center}
\end{figure}

\subsection{Comparison of Fourier {\rm [Fe/H]} with spectroscopic values}
\label{FevsSpec}

At this point, we would like to compare our photometric calculations
of the iron abundances with independent, solidly sustained values. We
have chosen to compare with the values given by \citet{Carretta2009} on
the spectroscopic scale [Fe/H]$_{\rm Carr}$, for the clusters in Table
\ref{MV_FEH:tab}. In Fig. \ref{PhoVSspec}, the photometric values
[Fe/H]$_{\rm UV}$ and [Fe/H]$_{\rm N}$, obtained as described in the
previous section, are plotted versus [Fe/H]$_{\rm Carr}$. Given that
the results for RRab and RRc stars come from different specific
calibrations, the comparison is done separately for both pulsation
modes. The top two panels, {\it a} and {\it b}, show the comparison of
[Fe/H]$_{\rm UV}$, obtained based on the calibrations of
eqs. \ref{eq:RR0_Fe} and \ref{eq:RR1_Fe}, duly transformed to the UVES
scale. These plots show good agreement between the photometric iron
abundance values obtained from the RRab and the RRc stars and the
spectroscopic values. There may be a slight hint that the RRc
calibration might need a small correction ($\sim -0.2$ dex) for the
most metal-poor clusters ([Fe/H]$_{\rm UV} \leq -1.8$ dex). The bottom
two panels, {\it c} and {\it d}, show the comparison of the values
[Fe/H]$_{\rm N}$, obtained from the calibrations of \citet{Nemec2013}
(eqs. \ref{NemAB} and \ref{NemC}). For the RRc star, the comparison is
satisfactory and comparable to the case in panel {\it b}. However, for
the RRab stars it is obvious from panel {\it c} that the metallicities
obtained from eq. \ref{NemAB} are largely overestimated. We do not
encourage the use of that calibration for the RRab stars. The same
conclusion was reached by Varga et al. in a poster delivered at this
IAU Symposium. See also the discussion at the end of the present
paper.

\subsection{Comparison of Fourier distances with {\sl Gaia} and {\sl HST} data based distances}
\label{distances}

\begin{figure}
\begin{center}
\includegraphics[scale=.8]{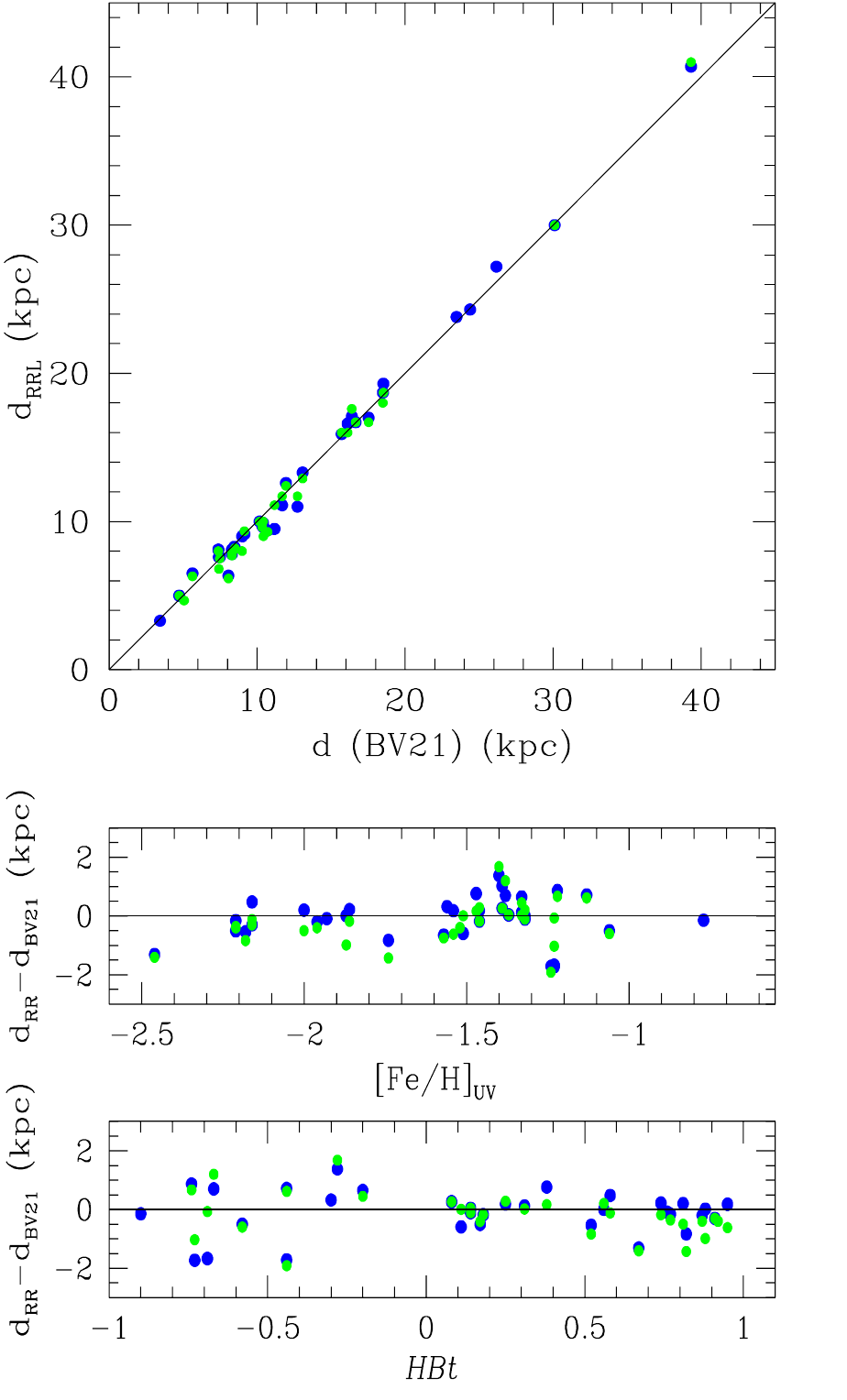}
\caption{Comparison of distances obtained from the RR Lyrae light
  curve Fourier decomposition and those of BV21. Blue and green
  symbols are for distances derived from the calibrations for RRab and
  RRc, respectively. All distance differences are contained within
  $\pm 1.9$ kpc of each other and the standard deviation is $0.7$
  kpc. (Updated from figure 4 in Arellano Ferro 2022, Rev. Mex. A\&A,
  58, 257). The color figure can be viewed online.}
\label{DISTANCIAS}
\end{center}
\end{figure}

We now proceed to compare the resulting distances from the absolute
magnitude determinations resulting from eqs. \ref{eq.RR0_Mv} and
\ref{eq.RR1_Mv} described in the previous section. The calculation of
the distance involves a value of $E(B-V)$; the adopted values are
listed in Table \ref{MV_FEH:tab}. We perform our comparison with
accurate critical mean distances from \citet[][BV21]{Baumgardt2021},
calculated for a large sample of globular clusters using data from
{\sl Gaia} early data release 3 (EDR3) parallaxes, line-of-sight
velocity dispersion profiles from {\sl Gaia} and {\sl HST}-based
proper motions. The top panel of Fig. \ref{DISTANCIAS} shows that our
distances compare very well with those of BV21. The distance
differences are always $< 1.9$ kpc and the standard deviation is $0.7$
kpc. The smaller bottom panels show the run of distance differences
with [Fe/H]$_{\rm UV}$ and with the horizontal branch (HB) structural
parameter $HBt$. In no case seems there to be a dependence of the
distances on either of these quantities.

Individual cluster distances are listed in Table \ref{T2:distance}. As
a complement we have estimated the cluster distances from the P--L
relation of SX Phe stars. We have taken into consideration the
relations derived by \citet[][AF11]{Arellano2011} and by
\citet[][CS12]{CohenSara2012}. A close look at the table corroborates that
these distances are consistent with the determinations from the RR
Lyrae stars calculated via the Fourier light curve decomposition.

\begin{table*}
\scriptsize
\begin{center}
\caption{Individual distances for a sample of globular clusters
  estimated homogeneously from the RR Lyrae stars' light curve Fourier
  decompositions and two versions of the SX Phe P--L relation.}
\label{T2:distance}

\begin{tabular}{lccccccc}
\hline
GC &$d (kpc)$&$d (kpc)$& $d$ (kpc) &No. of & $d$ (kpc)&$E(B-V)$ &d (kpc)\\
NGC(M)  & (RRab)& (RRc) & (SX Phe) &SX Phe&  (SX Phe)&mag & \\
   &  &   &P--L AF11 &&P--L CS12 & &BV21\\
\hline
 288  &9.0$\pm$0.2 &8.0 &8.8$\pm$0.4&6&9.4$\pm$0.6&0.03 & 8.988\\
1261 &17.1$\pm$0.4&17.6$\pm$0.7&--&--&--&0.01&16.400\\
1851 &12.6$\pm$0.2 &12.4$\pm$0.2&--&--& --&0.02&11.951\\
1904 (M79) &13.3$\pm$0.4&12.9&--& --&--&0.01& 13.078\\ 
3201 &5.0$\pm$0.2&5.0$\pm$0.1&4.9$\pm$0.3&16&5.2$\pm$0.4&dif& 4.737 \\ 
4147 &19.3 &18.7$\pm$0.5 &--& --&--&0.02&18.535\\   
4590 (M68)&9.9$\pm$0.3&10.0$\pm$0.2&9.8$\pm$0.5&6 &--&0.05& 10.404\\
5024 (M53)&18.7$\pm$0.4&18.0$\pm$0.5&18.7$\pm$0.6&13&20.0$\pm$0.8&0.02&18.498 \\ 
5053 &17.0$\pm$0.4 &16.7$\pm$0.4&17.1$\pm$1.1&12&17.7$\pm$1.2&0.02&17.537 \\
5272 &10.0$\pm$0.2&10.0$\pm$0.4&--&--&--&0.01&10.175\\
5466 &16.6$\pm$0.2 &16.0$\pm$0.6&15.4$\pm$1.3&5&16.4$\pm$1.3&0.00&16.120 \\
5904 (M5) &7.6$\pm$0.2 &7.5$\pm$0.3&6.7$\pm$0.5&3&7.5$\pm$0.2&0.03&7.479 \\
6139      &9.7$\pm$0.7 &9.6$\pm$0.6&-- &--&--&dif.&10.35 \\
6171 &6.5$\pm$0.3&6.3$\pm$0.2&--&--&--&0.33&5.631\\
6205 (M13) &7.6 &6.8$\pm$0.3&7.2$\pm$0.7&4&--&0.02&7.419 \\
6229 &30.0$\pm$1.5  &30.0$\pm$1.1 &27.9 & 1&28.9&0.01&30.106 \\
6254 (M10) &--&4.7&5.2$\pm$0.3&15& 5.6$\pm$0.3&0.25&5.067\\
6333 (M9) &8.1$\pm$0.2 &7.9$\pm$0.3&--&--& --&dif&8.300\\
6341 (M92)&8.2$\pm$0.2 &8.2$\pm$0.4&--&--& --&0.02&8.501\\
6362 &7.8$\pm$0.1&7.7$\pm$0.2&7.1$\pm$0.2&6&7.6$\pm$0.2&0.09&8.300\\
6366 &3.3 &--&--&--& --&0.80&3.444\\
6388 &9.5$\pm$1.2&11.1$\pm$1.1 &--&--&--&0.40&11.171\\ 
6401 &6.35$\pm$0.7 &6.15$\pm$1.4 &--&--&--&dif& 8.064\\
6402 (M14)&9.1$\pm$0.9 &9.3$\pm$0.5&--&--&--&0.57& 9.144\\
6441 &11.0$\pm$1.8 &11.7$\pm$1.0 &--&--& --&0.51&12.728\\ 
6712 &8.1$\pm$0.2&8.0$\pm$0.3&--&--&--&0.35&7.382\\
6779 (M56)&9.6&9.0& &--&--&0.26&10.430\\
6934 &15.9$\pm$0.4&16.0$\pm$0.6&15.8& 1&18.0&0.10&15.716 \\
6981 (M72)&16.7$\pm$0.4&16.7$\pm$0.4&16.8$\pm$1.6&3&18.0$\pm$1.0&0.06&16.661 \\
7006 &40.7$\pm$1.6&41.0$\pm$1.6&--&--&--&0.08&39.318\\
7078 (M15)&9.4$\pm$0.4&9.3$\pm$0.6&--& --&--&0.08& 10.709\\
7089 (M2) &11.1$\pm$0.6&11.7$\pm$0.02&--& --&--&0.06& 11.693\\
7099 (M30)&8.32$\pm$0.3 &8.1&8.0&1&8.3&0.03&8.458 \\
7492 &24.3&--&22.1$\pm$3.2&2& 24.1$\pm$3.7&0.00&24.390\\
Pal2 &27.2$\pm$1.8 &--&--&--&--&0.93& 26.174\\
Pal 13 &23.8$\pm$0.6&--&--&--&--&0.10&23.475\\
\hline
\end{tabular}
\end{center}
\end{table*}

\section{The $M_V$--[Fe/H] relation}
\label{MvFeRel}

Once we have proven that our photometric determinations of the iron
abundances and the absolute magnitudes (hence distances) compare
satisfactorily with sound and well-respected independent values, we
are in a position to revisit the $M_V$--[Fe/H] relation as implied by
our results. Before we carry on, let us recall that traditionally, the
$M_V$--[Fe/H] relation has been considered linear, of the form $M_V =
\mu $ [Fe/H] + $\gamma$, and that the values of the slope $\mu$ found
in the literature range between 0.09 and 0.30 depending on the authors
and approach \citep[e.g.][]{Demarque2000}.

\begin{figure}
\begin{center}
\includegraphics[scale=.7]{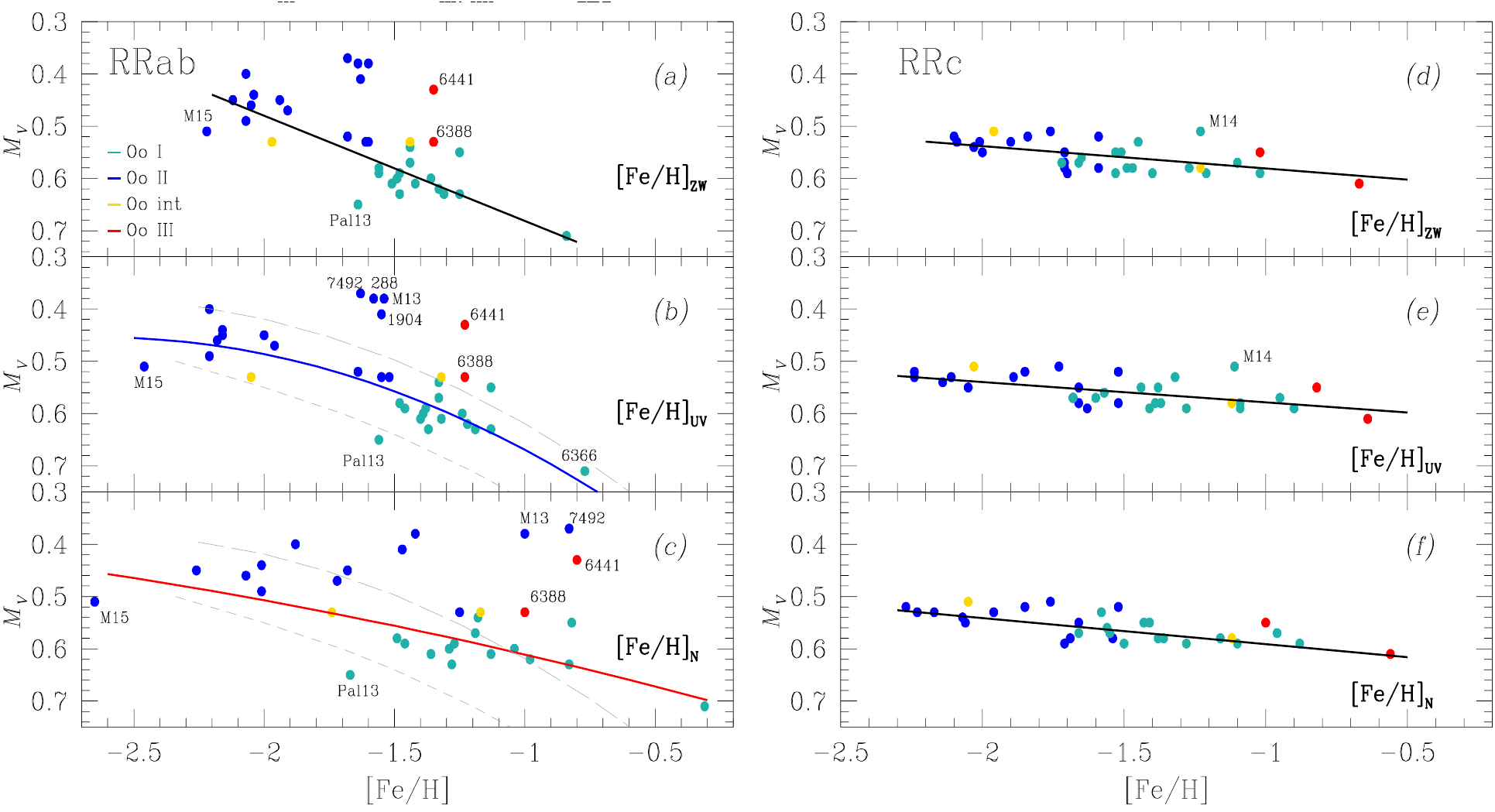}
\caption{The [Fe/H] versus $M_V$ correlations for RRab and RRc
  stars. The metallicity scales used are, from the top to bottom
  panels, [Fe/H]$_{\rm ZW}$, [Fe/H]$_{\rm UV}$ and [Fe/H]$_{\rm
    N}$. Cluster type is color coded in the legend in panel {\it
    (a)}. All fits have been weighted by the number of RR Lyrae stars
  included in each cluster. In the panel {\it (b)}, the gray curves
  are the theoretical predictions of \citet{Cassisi1999} (long dashes)
  and \citet{VandenBerg2000} (short dashes), which are remarkably similar to
  the photometric solution. The color figure can be viewed online.}
\label{MvFeH}
\end{center}
\end{figure}

We stress at this point that the values of $M_V$ and [Fe/H], obtained
from the calibrations for RRab and RRc stars, should not be mixed when
displaying the $M_V$--[Fe/H] relation since, in doing so, the
correlation will turn very noisy and become meaningless, as is obvious
from Fig. \ref{MvFeH} and the following discussion. In the left panels
of Fig. \ref{MvFeH}, we show the resulting $M_V$--[Fe/H] relation
based on the results for RRab stars. From top to bottom, panels {\it
  a}, {\it b} and {\it c}, we display the values of [Fe/H]$_{\rm ZW}$,
[Fe/H]$_{\rm UV}$ and [Fe/H]$_{\rm N}$, respectively, versus the
corresponding $M_V$ values from eq. \ref{eq.RR0_Mv}. Panel {\it a}
shows a linear run between [Fe/H]$_{\rm ZW}$ and $M_V$, but the
scatter is substantial. Interestingly enough, when [Fe/H]$_{\rm ZW}$
is converted to the spectroscopic scale, [Fe/H]$_{\rm UV}$, panel {\it
  b}, the correlation becomes less dispersed and, most remarkably, it
suggests a non-linear trend, very similar in fact to the theoretical
predictions of \citet{Cassisi1999} and \citet{VandenBerg2000}, shown as segmented
gray curves in the figure. To the best of our knowledge this is the
first empirical $M_V$--[Fe/H] relation that follows these theoretical
results, at least for RRab results. We will return to this point in
the Discussion. In panel {\it c} the metallicities from \citet{Nemec2013}
are considered. In this case, and due to the fact noted in Section
\ref{FevsSpec} that this calibration overestimates the metallicities,
the resulting $M_V$--[Fe/H]$_{\rm N}$ relation turns out to be very
scattered and with the RRab results systematically too large.

The quadratic fit in panel {\it b} was calculated while excluding the
obvious outliers (NGC 288, NGC 7492, M13 based on a single star, and
the OoIII type clusters NGC 6388 and NGC 6441), and including M15, and
weighing by the number of RRab stars considered in the calculation for
each cluster. The correlation is of the form,

\begin{eqnarray}
\label{FEHuv}
M_V= +0.081(\pm 0.060) \rm[Fe/H]_{\rm UV}^2 +0.428(\pm 0.207) \rm[Fe/H]_{\rm UV} +1.016(\pm 0.170),
\end{eqnarray}
\noindent
with r.m.s. = 0.060 mag. 

Let us turn to the case of the RRc stars, i.e., panels {\it d}, {\it
  e} and {\it f} on the right of Fig. \ref{MvFeH}. Now the three
panels display linear, tight and very similar correlations. The linear
fit in panel {\it e} is of the form,

\begin{eqnarray}
\label{FEHUVc}
M_V= 0.034 (\pm 0.009)\rm[Fe/H]_{\rm UV} + 0.601 (\pm0.015),
\end{eqnarray}
\noindent
with r.m.s. = 0.022 mag.

Note that the slope in the above correlation is shallower than the
shallowest $M_V$--[Fe/H] correlations found in the literature, but it
is tight and meaningful. It is also worth to stress that, for the RRc
stars, \citet{Nemec2013}'s calibration is not affected by the bias
noticed in their calibration of the RRab stars, and it produces [Fe/H]
values that are similar to those of the calibration of \citet{Morgan2007}
(panels {\it d} and {\it e}).

It shall be obvious now that mixing the RRab and RRc stars in a single
$M_V$--[Fe/H] relation will lead to a noisy and rather meaningless
relation. While this requires some explanation, we shall defer our
speculations to the Conclusions.

\section{The role of the HB structure parameter}
\label{HBparam}

In a theoretical study, \citet{Demarque2000} concluded that the
$M_V$--[Fe/H] relation is anything but universal, that the slope is
also a function of the metallicity range considered, and that for a
given metallicity, the luminosity depends on the HB
morphology. Following \citet{Demarque2000}, we have adopted the HB type
parameter defined as $HBt$ $\equiv (B-R) / (B+V+R)$ \citep{Lee1994},
where $B, V$ and $R$ are the numbers of HB stars to the blue of the
instability strip, RR Lyrae, and those to the red of the instability
strip, respectively. We have estimated the value of $HBt$ from the
resulting color--magnitude diagrams (CMDs) for almost all clusters in
Table \ref{MV_FEH:tab}. Before counting stars we performed a
membership analysis using the proper motions, radial velocities and
positions available in the different {\sl Gaia} data releases;
\citep{Gaia2016} and Gaia collaboration (2022) and the membership analysis approach of
\cite{Bustos2019}. For a few clusters for which we have no CMD, we
adopted $HBt$ from \citet{Torelli2019}, who also applied membership
considerations.

In Fig. \ref{LMvFe} the well-known dependence of $HBt$ on metallicity
is shown in the left panels for the RRab and the RRc iron abundance
solutions. In the right panels, the run of the absolute magnitude
$M_V$ with $HBt$ is displayed, again separately for the two pulsation
modes. The striking feature of these plots is the quadratic
correlation between $M_V$ and $HBt$ for the RRab solutions, plus the
obvious linear correlation of these two quantities for the RRc
solutions. The fact that the three parameters [Fe/H]$_{\rm UV}$, $M_V$
and $HBt$ are interrelated is clear. A multiple regression analysis
leads to the following correlations for the RRab and the RRc stars,
respectively:

\begin{figure}
\begin{center}
\includegraphics[scale=.7]{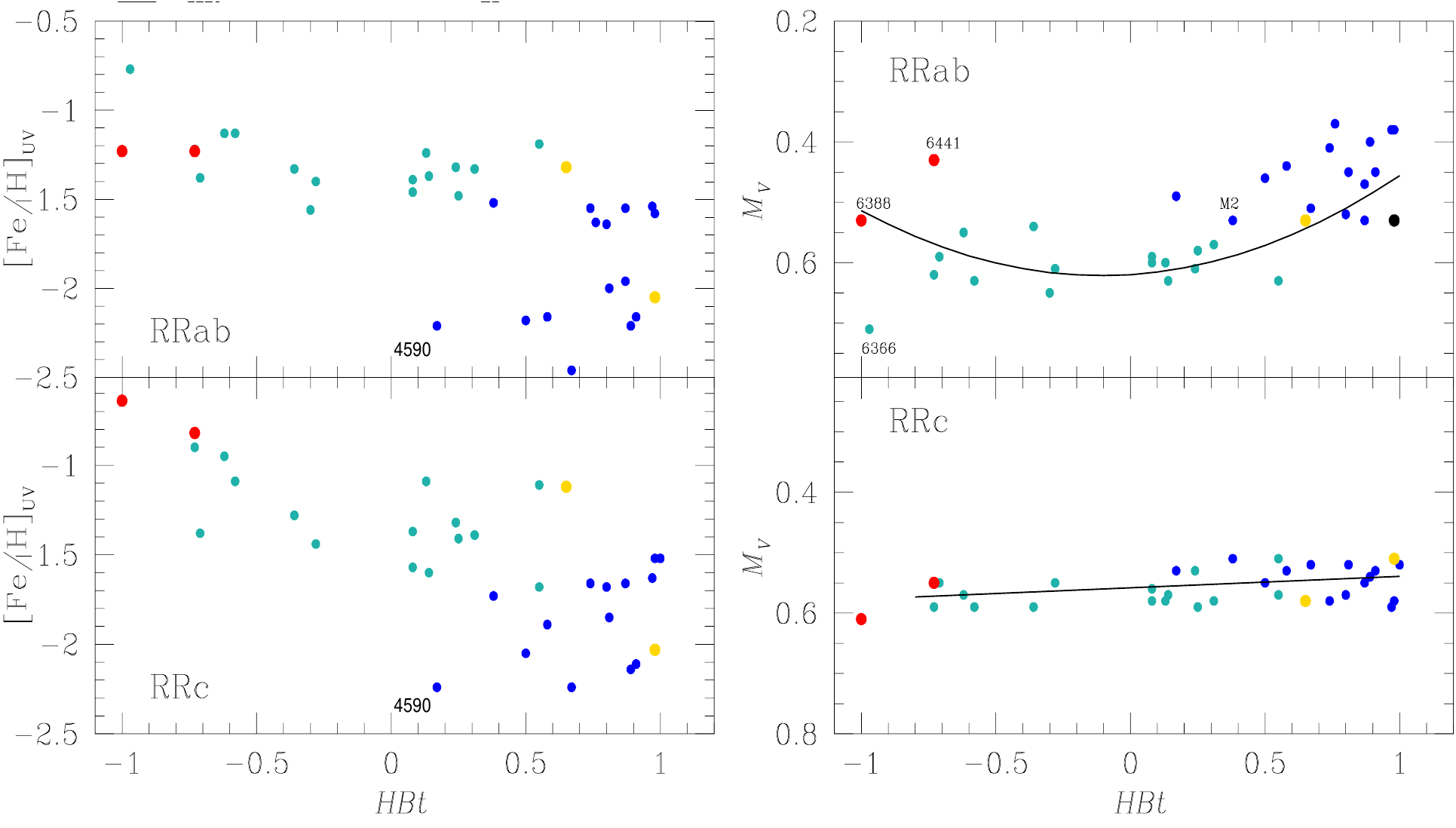}
\caption{Interdependencies of $M_V$, [Fe/H]$_{\rm UV}$ and the $HBt$
  parameter from the solutions for the RRab and the RRc stars. The
  color figure can be viewed online.}
\label{LMvFe}
\end{center}
\end{figure}

\begin{eqnarray}
\label{MVFELab}
M_V=  A +  B~ {\rm[Fe/H]}_{\rm UV} + C~ {\rm[Fe/H]}_{\rm UV}^2 + D~{\it HBt} + E~{\it HBt}^2,
\end{eqnarray}
with $A=+1.096 (\pm 0.141), B=+0.519 (\pm 0.172), C=+0.119 (\pm
0.050), D=+0.006 (\pm 0.014), E=-0.111 (\pm 0.029)$, and r.m.s. =
0.053 mag.

\begin{eqnarray}
\label{MVFELc}
M_V= +0.609 (\pm 0.016) + 0.032 (\pm 0.009)~ \rm[Fe/H]_{\rm UV} + 0.015 (\pm 0.011)~{\it HBt},
\end{eqnarray}
with  r.m.s. = 0.024 mag.

We note that in eq. \ref{MVFELab} the coefficient $D$ is not
significant and that term can therefore be neglected. However, $E$ is
significant, hence the quadratic dependence of the calibration on
$HBt$ as well as on [Fe/H]$_{\rm UV}$. In eq. \ref{MVFELc} the
coefficient of $HBt$ is not significant, implying a linear correlation
between $M_V$ and [Fe/H]$_{\rm UV}$ for the RRc stars' solutions. If
this latter term is ignored, we can see that eqs. \ref{MVFELc} and
\ref{FEHUVc} are virtually identical.

The results exposed in this section are in a way the empirical
confirmation of the theoretical argumentation of \citet{Demarque2000},
i.e., that a simple linear $M_V$--[Fe/H] relation for RR Lyrae in
globular clusters is not sufficient but that, for a given metallicity,
the luminosity depends on the HB structure, which can be described by
the $HBt$ parameter. To this statement we add that this seems indeed
the case when the HB luminosities are calculated from RRab stars,
whereas the relation remains simple and linear if the luminosity
indicators are the overtone RRc stars. If the analysis is performed
without distinguishing between both pulsation modes, the presence of
RRab stars will contribute to the scatter and non-linearity between
$M_V$ and [Fe/H]. If RR Lyrae stars are to be used as distance
indicators, one should be inclined to prefer the simple RRc stars, and
a proper instrument for this purpose could be the $M_V$--[Fe/H]
relation offered in this work in the form of eq. \ref{FEHUVc}.

\section{Variable Stars in our sample of Globular Clusters}
\label{VarsGC}

The exercise of obtaining accurate photometry in the field of our
sample of globular clusters via the DIA, for a time-series of images,
as described in Section \ref{DIA}, produces light curves of many point
sources in the field of view. Typically between 1000 and 15,000 stars
can be isolated and measured, depending on the cluster distance, size,
reddening and the prevailing sky conditions during our
observations. Exploring all light curves, employing different methods,
became a parallel routine in the project. This allowed us to recover
the light curves of all known variables and, very often, to find
variables not detected previously. The analysis of the light curve
shape, period, and the position of the star in the corresponding CMD,
generally enables a proper classification of the variable type.

A noteworthy example of the discovery of unexpected variable stars is
the case of Palomar 2, well known since a long time as a cluster
devoid of variables. However a detailed analysis of our data led to
the discovery of 20 RRab stars and 1 RRc. A revision of {\sl Gaia} DR3
data enabled us to identify 10 additional variables. It has been shown
that 16 RRab, 1 RRc and 1 RGB are cluster member stars \citep{Arellano2023}.

Table \ref{NoVARS} summarizes the number of variables and their types
found during the development of our project. All pertinent details for
each cluster, such as their light curves, ephemerides, Fourier fits,
classifications and other discussions as to cluster membership and
specific peculiarities, can be found in the many original papers
referenced and properly coded in table 3 of \citet{Arellano2022}. The present
Table \ref{NoVARS} has been updated and edited, and it supersedes
table 3 of \citet{Arellano2022}. In Table \ref{NoVARS} we have not included any
variable star listed as such in the {\sl Gaia} DR3 in the
corresponding field of each cluster. Many of the {\sl Gaia} variables
coincide with previously known variables and/or may be field
stars. Therefore, while we are not aware of a dedicated analysis of
these variables, as for example for the case of Palomar 2 \citep{Arellano2023}
or NGC 6139 (Yepez et al. 2023), we have based our star counts on the
present edition of the Catalog of Variable Stars in Globular Clusters
\citep{Clement2001} and our own findings.

\begin{table*}
\scriptsize
\begin{center}
\caption{Number of presently known variables per type and per cluster
  for the most common variable types, in the globular clusters studied
  by our group$\dagger$.}
\label{NoVARS}
\begin{tabular}{lccccccccc}
\hline
GC & RRab & RRc &RRd & SX Phe& Binaries& CW-(AC)-RV& SR, L, M& unclass& Total per cluster\\
NGC (M) & & & & & & & &others $*$& \\
\hline
288  &0/1&0/1&0/0&0/6 &0/1&0/0&0/1&--&0/10\\
1261 &0/16 & 0/6 & 0/0 & 0/3 & 0/1 & 0/0 & 0/3 &--& 0/29 \\
1904 (M79)&0/6&1/5&0/0&0/5&0/1&0/1&0/14&0/1&1/32\\ 
3201 &0/72&0/7&0/0&3/24&0/11&0/0&0/8&0/7&3/122\\ 
4147 &0/5 &0/19 &0/1 &0/0 &0/14 &0/0 &2/2 &0/3&2/41\\  
4590 (M68)&0/14 &0/16&0/12&4/6&0/0&0/0&0/0&1/2&4/48\\
5024 (M53)&0/29&2/35&0/0&13/28&0/0&0/0&1/12&--&16/104\\ 
5053 &0/6 &0/4&0/0&0/5&0/0&0/0 &0/0&--&0/15\\
5466 &0/13 &0/8&0/0&0/9&0/3&0/1&0/0&2/2&0/34\\
5904 (M5)&2/  89&1/40&0/0&1/6&1/3&0/2 &11/12 &0/1&16/152\\
6171 (M107)&0/15 &0/6 &0/0 &0/1 &0/0 &0/0 &2/3 &0/3&2/25\\
6139       &0/4 &4/5 &0/0 &0/0 &1/1 & 0/0 &4/7 &   &9/17\\
6205 (M13)& 0/1 & 1/7& 2/2 & 2/6 & 1/3 & 0/3 &3/22 &0/4 & 9/44\\
6229 &10/42 &5/15 &0/0 &1/1&0/0 &2/5&6/6 &0/1&24/69\\
6254 (M10)&0/0 & 0/1 &0/0 & 1/15 & 2/10 & 0/3 & 0/5 &0/2 & 3/34 \\
6333 (M9)&0/8 &2/10&1/1&0/0&3/4&1/1&5/6&3/4&12/30\\
6341 (M92)&0/9 &0/5 & 1/1 & 1/6 & 0/0 & 0/1 & 1/1 &0/6 & 3/23\\
6362 &0/16 &0/15 &1/3& 0/6 & 0/12 & 0/0&0/0 &0/22& 1/52 \\
6366 &0/1 &0/0&0/0&1/1 &1/1&1/1&3/4&--&6/8\\
6388 &1/14&2/23&0/0&0/1&0/10 &1/11&42/58&--&46/117\\
6397 &0/0 &0/0&0/0&0/5 &0/15&0/0&0/1&0/13&0/21\\
6401 &6/23&6/11&0/0&0/0&0/14&0/1&3/3&14/14&15/52\\
6402 (M14)&0/55 &3/56&1/1&1/1 &0/3& 0/6& 18/32& &23/154\\
6441&2/50 &0/28&0/1& 0/0&0/17&2/9&43/82&0/10&47/187\\
6528&1/1&1/1&0/0&0/0&1/1&0/0&4/4&--&7/7\\
6638&3/10&2/18&0/0&0/0&0/0&0/0&3/9&0/25&8/37\\
6652&0/3&0/1&0/0&0/0&1/2&0/1&0/2&1/5&1/9\\
6712& 0/10 & 0/4 & 0/0 & 0/0 & 2/2 &0/0& 5/11 &0/8  &7/27 \\
6779 (M56)& 0/1 &0/2&0/0&1/1 &3/3&0/2&0/3&1/6&4/12\\
6934& 3/68 & 0/12 & 0/0 & 3/4 &0/0 & 2/3 &3/5 &1/6&11/92\\
6981 (M72)&8/37 &3/7&0/0&3/3&0/0&0/0&0/1& &14/48\\
7078 (M15)&0/65 &0/67&0/32&0/4&0/3&0/2&0/3&0/11&0/176\\
7089 (M2)&5/23 &3/15&0/0&0/2&0/0&0/4&0/0&0/12&8/44\\
7099 (M30)&1/4&2/2&0/0&2/2&1/6&0/0&0/0&0/3& 6/14\\
7492 &0/1  &0/2&0/0&2/2&0/0&0/0&1/2&--& 3/7\\
Pal 2&16/16&1/1& 0/0 & 0/0& 0/0 & 0/0 & 1/1 &--& 18/18 \\
Pal 13 &0/4 & 0/0 &0/0 & 0/0 & 0/0& 0/0 & 1/1 & & 1/5 \\
\hline
Total per type &58/732 &41/454 &6/54 &39/153 &17/141 &9/57 &162/324&{\it 23/171} * &330/1916\\
\hline
\end{tabular}
\small
\center{\quad $\dagger$. The variable star types have been adopted
  from the General Catalog of Variable Stars
  \citep{Kazarovets2009,Samus2009}. Entries expressed as $M/N$ indicate
  the $M$ variables found or reclassified by our program and the total
  number $N$ of presently known variables. Relevant papers on
  individual clusters can be found properly coded in table 3 of
  \citet{Arellano2022}.

\quad $*$. Numbers from this column are not considered in the totals,
since they include unclassified or likely field variables.\\ }
\end{center}
\end{table*}

\section{Summary and Conclusions}
\label{Summary}

For nearly two decades, systematic observational scrutiny of the
variable star populations in a sample of 39 globular clusters by means
of time-series CCD imaging in the \emph{VI} passbands, and their
treatment by the DIA, enabled to collect numerous light curves of
variable stars, of many types, contained in these stellar systems. Of
particular interest to the present work were the light curves of RR
Lyrae stars of both pulsation modes, the fundamental RRab and the
first overtone RRc groups. We have performed Fourier decompositions of
these light curves and the resulting Fourier parameters have been
employed in carefully selected empirical calibrations to calculate, in
an unprecedentedly homogeneous manner, the atmospheric iron
abundances, [Fe/H], and absolute magnitudess, $M_V$, via this
photometric approach. Attention was paid to the zero points of the
absolute magnitude calibrations.

The mean $M_V$ values and estimates of the interstellar reddening led
to photometric distances calculated for our sample 39 globular
clusters and reported here. The mean iron abundance for a given
cluster was calculated on the scale of \citet{Zinn1984} [Fe/H]$_{\rm
  ZW}$, transformed to the UVES spectroscopic scale of
\citet{Carretta2009} [Fe/H]$_{\rm UV}$, and on the scale of
\citet{Nemec2013} [Fe/H]$_{\rm N}$.

The photometric [Fe/H]$_{\rm UV}$ and the spectroscopic [Fe/H]$_{\rm
  CARR}$ values taken from \citet{Carretta2009} were found to be in very
good agreement for both the solutions from the RRab and the RRc
stars. \citet{Nemec2013}'s calibration for the RRc stars are also in good
agreement with both [Fe/H]$_{\rm UV}$ and [Fe/H]$_{\rm CARR}$. It was
found, however, that the photometric values from the calibration for
RRab stars of \citet{Nemec2013} are systematically overestimated relative
to the spectroscopic values. This result becomes of particular
relevance since the calibration of \citet{Nemec2013} was used by {\sl
  Gaia} to calculate the iron abundances from the photometry of RR
Lyrae stars. It is worth noting that a similar result has been
reported in a poster presented at this meeting by Varga et al. See
also the Q\&A section below for further comments on this issue.

Similarly, we have compared our photometric determinations of the
accurate mean cluster distances derived by \citet{Baumgardt2021}. The
agreement is outstanding, yielding distance differences within $\pm
1.9$ kpc for the entire distance range up to 40 kpc.

The resulting $M_V$--[Fe/H] relation for the RRab stars looks
non-linear and scattered. However, based on the RRc stars the relation
remains linear and tight. The latter relation shows only a mild but
significant dependence of the luminosity on the metallicity. It is
clear that if the $M_V$ and [Fe/H] results from RRab and RRc stars are
mixed, the resulting $M_V$--[Fe/H] relation will be apparently linear,
very scattered and steeper. We find that it is important to segregate
the RRab and RRc results for a more meaningful interpretation of the
$M_V$--[Fe/H] relation.

As suggested on theoretical grounds \citep{Demarque2000}, we have
empirically shown that the HB type parameter, $HBt$, plays a
significant role in the $M_V$--[Fe/H] relation if calculated using the
RRab stars as tracers, in which case the relation becomes quadratic in
[Fe/H] and $HBt$. No dependence on $HBt$ was found for the RRc stars.
These results strengthen the notion that RRc stars may be more trusted
distance indicators than their more complex RRab counterparts.

While I really do not have an in-depth explanation for the differences
in the luminosity dependence on the metallicity for RRab and from RRc
stars, it may be relevant to mention that RRab stars carry a greater
degree of complexity; they have larger amplitudes, many of them have
(undetected) Blazhko modulations, their light curves are asymmetrical
and hence prone to more inaccurate Fourier decompositions; as their
periods are often longer, their light curves are not fully or rather
unevenly covered. Their evolutionary tracks toward the RGB are spread
more widely with small variations in age, inner mass structure and
chemical differences. The RRc stars, on the other hand, are simpler,
easier to decompose and their luminosities are less sensitive to age,
mass or chemistry. However I emphasize that the remarks in the present
work may require further theoretical input.

Systematic time-series CCD imaging and differential image analysis
performed over the last nearly 20 years has enabled us to report and
classify 330 newly detected variables in our sample of globular
clusters.

\section{Acknowledgments}

I am indebted to all my friends and colleagues who have accompanied me
on this globular cluster journey that has lasted nearly two
decades. They number too many to specifically name them all here;
their names are all included as co-authors of our papers; to all of
them, my gratitude for having shared with me and taught me their
expertise. I am particularly thankful to the Time Allocation
Committees of several observatories for having generously granted the
time needed to push forward our observing strategies. For many years,
the project has received support from the Institute of Astronomy and
DGAPA, of the National Autonomous University of Mexico, through a
range of grant numbers.

\printbibliography

\vspace{1.0cm}

{\bf \large Questions and Answers}
\vspace{1.0cm}

{\bf Clara Mart\'inez-V\'azquez}: Thanks for your talk and the
interesting comments offered! I would like to bring to your attention
a set of recently published new $P$--$\phi_{31}$--[Fe/H] relationships
for RRab and RRc types. They have been calibrated based on a very
large and homogeneous sample of spectroscopic metallicities. The
testing of the relations has proven to be very efficient. Here are the
papers in case you want to check them:
https://iopscience.iop.org/article/10.3847/1538-4357/abefd4
($P$--$\phi_{31}$--[Fe/H] for RRab) and
https://iopscience.iop.org/article/10.3847/1538-4357/ac67ee
($P$--$\phi_{31}$--[Fe/H] for RRc).

{\bf AAF}: Thank you so much for the references. I will check them
carefully.
\vspace{0.4cm}

{\bf H\'ector V\'azquez Raml\'o}: Thanks for the interesting talk. A
technical question: if I understood you correctly, you use
differential imaging to obtain the light curves of the RR Lyrae in
globular clusters. That implies soundly coping with PSF variations
from image to image (at least if observations were obtained from the
ground, I didn't get if that is the case). Could you expand a little
on this, please? Which package do you use?

{\bf AAF}: We use Difference Image Analysis (DIA) through the pipeline
DanDIA (Bramich 2008, MNRAS, 386, L77; Bramich et al. 2013 MNRAS, 428,
2275). DanDIA is constructed based on the DanIDL library of IDL
routines available at http://www.danidl.co.uk.  Our use is described
in detail in Bramich et al. (2011 MNRAS, 413, 1275). A recent paper
where you can gain an overall view of all our procedures is perhaps
Yepez et al. (2020, MNRAS, 494, 3212). Regarding the PSF, DanDIA
constructs a reference image from a combination of the best images in
our collection. Then, it isolates up to 400 stars in the image using
some isolation criteria, and it uses those individual PSFs to search
for possible trends across the chip, fitting high-order
polynomials. The PSF kernel is then used for the individual
differential images.
\vspace{0.4cm}

{\bf H\'ector V\'azquez Raml\'o}: Another question regarding the
statement of RRc being more suitable for distance measurements, that
was an interesting one. It is clear that RRab are more widely used due
to their much higher abundance and their higher pulsation amplitude
(easier to identify) in comparison with RRc. Out of curiosity, I
really don't know: are there studies exclusively focusing on RRc for
estimating distances? Is there a general consensus in the community on
that issue?

{\bf AAF}: There are of course papers dedicated exclusively to RRc
stars, for instance one on spectral metallicity determinations by
Sneden et al., (2018, AJ, 155, 45), but not on the premise that they
may be better distance indicators, I think. This is, in fact, one of
the highlights among our results; this, and the fact that absolute
magnitudes of mixed samples of RRab and RRc stars will lead to larger
uncertainties and mislead calibrations of the famous $M_V$--[Fe/H]
correlation. I do not think there is, at present, a general consensus
on these points, or not just yet. The community has become more
interested in the presence of secondary frequencies caused by
non-radial modes in RRc stars, I believe, the ones with ratios relative
to the overtone frequency ($f_1$) known as the $f_{61}/f_1 \sim
0.61$--$0.65$.
\vspace{0.4cm}

{\bf Gergely Hajdu}: Thank you very much for the update on globular
cluster RR Lyrae. I would like to discourage the use of the Nemec et
al. (2013) relations for metallicity estimation of RRab stars. They
come from a very small sample of variables, and the authors overfit
their data with parameters that are highly correlated (especially
their three parameters that all contain $\phi_{31}$), as shown by the
large formal errors (4.64 dex on the intercept!). Unfortunately, the
baseline {\sl Gaia} DR3 [Fe/H] estimates are based on this formula,
leading to clearly artificial features appearing when the Bailey
diagram is colored according to that metallicity estimate (i.e.,
fig. 35 of
https://ui.adsabs.harvard.edu/abs/2022arXiv220606278C/abstract).

{\bf AAF}: Thanks for your comments. As you may have noticed, in our
comparison of the photometric estimates of [Fe/H]$_{\rm UV}$ vs. the
spectroscopic values, [Fe/H]$_{\rm CARR}$, we also noticed that
something does not work well for the calibration of Nemec et
al. (2013) for the RRab stars, since their calibration clearly
overestimates the iron abundance values. For the RRc stars, it seems
as good as the calibration of Morgan et al. (2007). Thus, it is
interesting to note your comments and recommendation against that
calibration. By the way, there is a poster at this Symposium by Varga
et al.; the authors reached the same conclusion as regards Nemec et
al.'s calibration for the RRab and, apparently, they have corrected
it.
\vspace{0.4cm}

{\bf V\'azsony Varga}: I also recommend comparision with Fabrizio et
al.'s (2021) $\Delta S$ Bailey diagram metallicity distribution. I'd
also like to add that the anomalous metal-rich regions in the current
{\sl Gaia} Bailey diagram are likely caused by the small parameter
space covered by the calibration data set.

{\bf AAF}: Thank you for your comments. I have noticed that you also
highlight that the calibration of \citet{Nemec2013} overestimates the
iron abundances for RRab stars.
\end{document}